\documentclass[manuscript,superscriptaddress,floats,prb,floatfix,nobibnotes]{revtex4}
\usepackage{amsmath}
\usepackage{amssymb}
\usepackage{bm}
\usepackage{graphicx,epsfig,color}

\begin{document}

\title{Correlated interaction fluctuations in photosynthetic complexes}

\author{Sebastiaan M. Vlaming}\email{vlaming@mit.edu}
\affiliation{Center for Excitonics and Department of Chemistry, Massachusetts Institute of Technology,
77 Massachusetts Avenue, Cambridge, Massachusetts 02139, United States}
\author{Robert J. Silbey}\thanks{deceased}
\affiliation{Center for Excitonics and Department of Chemistry, Massachusetts Institute of Technology,
77 Massachusetts Avenue, Cambridge, Massachusetts 02139, United States}

\date{\today}

\begin{abstract}
The functioning and efficiency of natural photosynthetic complexes is strongly influenced by their embedding in a noisy protein environment, which can even serve to enhance the transport efficiency. Interactions with the environment induce fluctuations of the transition energies of and interactions between the chlorophyll molecules, and due to the fact that different fluctuations will partially be caused by the same environmental factors, correlations between the various fluctuations will occur. We argue that fluctuations of the interactions should in general not be neglected, as these have a considerable impact on population transfer rates, decoherence rates and the efficiency of photosynthetic complexes. Furthermore, while correlations between transition energy fluctuations have been studied, we provide the first quantitative study of the effect of correlations between interaction fluctuations and transition energy fluctuations, and of correlations between the various interaction fluctuations. It is shown that these additional correlations typically lead to changes in interchromophore transfer rates, population oscillations and can lead to a limited enhancement of the light harvesting efficiency.
\end{abstract}

\maketitle

\section{Introduction}

The harvesting of sunlight by organisms such as plants and bacteria plays a crucial role in life on earth. Nature has optimized the photosynthesis process to a great extent, leading to a high efficiency in both the capturing of light and the subsequent energy transport within the organism's light harvesting complexes.~\cite{Amerongen00, Sundstrom99, Sumi99, Mukai99, McDermott95} Pigment molecules in these light harvesting structures absorb the light, after which the excitation energy is eventually funneled towards the reaction center, where charge separation takes place and the energy can be utilized for biochemical purposes. A closer understanding of the functioning of these biological systems and what features are crucial to their high efficiency may provide further insight in how to optimize synthetic light harvesting structures.

Recent experimental studies have shown that quantum coherence in light harvesting complexes persists on surprisingly long timescales.~\cite{Brixner05,Engel07, Lee07, Panit10} While excitation energy transport (EET) in such molecular systems has been a subject of study for decades,~\cite{Forster48, Grover70, Kenkre74} the aforementioned experimental observation of long-lived coherence despite its occurrence in noisy environments at room temperature has spurred further theoretical research. Naively, one might think that coupling the electronic excitations to a dissipative environment will lead to rapid decoherence and a loss of excitation energy. However, it has been shown that environment-induced dephasing and noise may actually enhance the transport efficiency of excitations in such structures.~\cite{Gaab04, Mohseni08, Plenio08, Ishizaki09, Rebentrost09, Wu10} The rationale behind this lies in the fact that interaction with degrees of freedom in the bath can induce transitions of the excitation that are directed towards the reaction center. Recent studies have shed further light on the mechanisms that allow for optimal efficiency of the excitation, suggesting an intricate functioning where aspects such as the non-Markovian nature of the bath interactions~\cite{Rebentrost09-2, Roden09, Chen11} and correlations in the bath~\cite{Adolphs06, Caruso09, Wu10, Strumpfer11} may be exploited to achieve further optimization of the light harvesting complex.

 The interaction of the chromophores with their local environments leads to fluctuations of the transition energies and interactions. More specifically, changes in the environment will induce fluctuating local electrical fields. The resultant Stark shifts in the transition energies and transition dipoles of the chromophores will thus also fluctuate; in addition, environment-induced changes in relative position or orientation of the chromophores will also cause corresponding fluctuations in the interactions.~\cite{Adolphs06} A number of studies have been performed where only the effect of transition energy fluctuations has been included, as well as possible spatial and temporal correlations between these transition energy fluctuations.~\cite{Adolphs06, Caruso09, Wu10, Strumpfer11} This is commonly done in the framework of the extended Haken-Strobl-Reineker model,~\cite{Haken72, Haken73} which provides an attractive approach due to its simplicity and tractability. The original Haken-Strobl-Reineker model allows for a treatment of both transition energy and interaction fluctuations, and while it does not include correlations and inherently assumes a high temperature, the formalism can straightforwardly be generalized to remove the former drawback. Fluctuations in the interactions, in contrast to transition energy fluctuations, have typically not been accounted for in studies of photosynthetic complexes, despite the fact that these will generally occur as well.

 Fluctuations in transition energies and interactions are induced by variations in the local environments of the chromophores, and environmental variations around one chromophore will generate both fluctuations in its transition energy and in its interaction with the other chromophores. In addition, different chromophores may share part of the environment; both these arguments imply that the generated fluctuations will in general be correlated. While correlations between transition energy fluctuations have been studied,~\cite{Adolphs06, Caruso09, Wu10, Strumpfer11} there should generally also be correlations of transition energy fluctuations and interaction fluctuations, and between different interaction fluctuations. In previous studies, such correlations have commonly been discarded, even though there is no a priori reason to do so. A number of recent studies have also suggested the occurrence of such fluctuations in photosynthetic complexes.~\cite{Ulrich11, Shim11, Coker11} Moreover, fluctuations of the various intermolecular interactions are usually not accounted for, although also these are typically induced by changes in the local environment. We present a study of the effects of fluctuations of the interactions and their possible correlations with the various transition energy and other interaction fluctuations, and provide a numerical analysis of their measurable effects. When applied to natural photosynthetic complexes, such as the Fenna-Matthews-Olson (FMO) complex,~\cite{Fenna75, Olson80, Blankenship88, Li97, Vulto98} it is shown that interaction fluctuations can have a considerable impact on the excitation dynamics and the efficiency, and these effects should generally not be neglected. Furthermore, we show that additional correlations can lead to enhanced oscillations of the exciton populations and modifications of the transfer and dephasing rates, and we quantify the dependence of the trapping efficiency on the correlations and initial conditions.

The paper is structured as follows. In Sec. \ref{Sec:Theory}, we provide the theoretical background of our study: in Sec. \ref{Sec:HSR}, we introduce the Haken-Strobl-Reineker formalism that is used to describe the effect of environmentally induced fluctuations and their possible correlations on the excitation dynamics, Sec. \ref{Sec:corr} discusses the possible correlations, and in Sec. \ref{Sec:FMO} we introduce the Fenna-Matthews-Olson photosynthetic complex that we will apply our theory to. In Sec. \ref{Sec: numerical}, we show the results of application of our theory to FMO: Sec. \ref{Sec:uncorrelated fluctuations} elucidates the role of uncorrelated interaction fluctuations, Sec. \ref{Sec:EJfluc} focuses on the effect of correlations between fluctuations in transition energies and interactions, while subsequently in Sec. \ref{Sec:JJfluc} we investigate the effect of correlated interaction fluctuations. Our conclusions are presented in Sec. \ref{Sec:conclusions}.

\section{Theory}
\label{Sec:Theory}

\subsection{The Haken-Strobl-Reineker model}\label{Sec:HSR}

Upon absorption of the solar light and energy transfer into the FMO complex, an electronic excited state will be created in the bacteriochlorophylls (BChl's). Due to the strong interchromophore interactions, this excited state (exciton) will be delocalized over a number of BChl's. To describe the excitation and its dynamics, we employ the Frenkel exciton Hamiltonian,~\cite{Davydov71, Agranovich82}

\begin{equation}H=H_S+H_R+H_{S-R}=\sum_nE_n \left|n\right>\left<n\right|+\sum_{n,m\neq n} J_{nm}\left|n\right>\left<m\right|+\sum_{q}\omega_qb_q^{\dagger}b_q+H_{S-R},\end{equation}

where $\left|n\right>$ describes a state where chromophore $n$ is in its excited state while all others are in their ground states, $J_{nm}$ is the interaction between chromophores $n$ and $m$, and we have a bath of harmonic modes labeled by $q$. The first two terms are the system Hamiltonian $H_S$, the third term corresponds to the bath Hamiltonian $H_R$, and the final term describes the coupling between the system and the bath $H_{S-R}$, which we will keep unspecified for now. In our case, the system consists of the BChl's within the photosynthetic complex, while the environment includes any other degrees of freedom the excitations can couple to, such as the protein surroundings. To adequately describe the environmental effects, it is necessary to work in the density matrix formalism.~\cite{Neumann27, Blum81, May00} The density matrix corresponding to the wave function $\left|\psi(t)\right>$ is given by $\tilde{\rho}(t)=\left|\psi(t)\right>\left<\psi(t)\right|$, and evolves according to the Liouville-von Neumann equation ($\hbar$ has been set to unity), \begin{equation}\label{Liouville} i \frac{\partial \tilde{\rho}(t)}{\partial t}=\left[H(t),\tilde{\rho}(t)\right]\equiv \mathcal{L}\tilde{\rho}(t).\end{equation}
Here, we have defined the Liouville superoperator as $\mathcal{L}=\left[H,...\right]$. The above time evolution equation of the density matrix is equivalent to the time evolution of the wave function $\left|\psi(t)\right>$ as given by the Schr\"odinger equation. At this point, we switch to the interaction picture, where the time evolution induced by the Hamiltonian terms $H_S+H_R$, which are large compared to the system-bath interaction $H_{S-R}$, is explicitly removed from the time evolution of all operators $A(t)$, \begin{equation}A_I(t)=e^{i\left(H_S+H_R\right)t}A(t)e^{-i\left(H_S+H_R\right)t}.\end{equation}
We are primarily interested in the evolution of the system degrees of freedom, which we obtain by taking the trace over the bath degrees of freedom. This leaves us with the reduced density matrix $\rho(t)$ in the interaction picture, which is the quantity we will consider from this point on when referring to the density matrix.

The approach we take in treating interactions with the environment is based on a method first introduced by Haken, Strobl and Reineker.\cite{Haken72, Haken73} In this approach, one models the bath-induced fluctuations as classical Gaussian Markov processes. While the original Haken-Strobl-Reineker (HSR) model assumed uncorrelated site energy fluctuations, one can extend the methodology to allow for correlations between the various fluctuations; we refer to Ref. \onlinecite{Chen10} for the details. The Hamiltonian can then be written as \begin{equation}\label{HamHSR}H=\sum_nE_n \left|n\right>\left<n\right|+\sum_{n,m\neq n} J_{nm}\left|n\right>\left<m\right|+\sum_{n,m}V_{nm}(t)\left|n\right>\left<m\right|.\end{equation} The terms $V_{nm}(t)$ are the environmentally induced fluctuations of the various system Hamiltonian matrix elements. Note that the Hermiticity of the Hamiltonian implies $V_{nm}(t)=V_{mn}^{*}(t)$. The averages $\left<V_{nm}(t)\right>$ can be set to zero without loss of generality, and since the fluctuations are assumed to be Gaussian Markov processes, the problem is fully defined when the correlation functions $\left<V_{nm}(t)V_{n'm'}(t')\right>$ are known. The effect of the environment is now fully encoded in the bath correlation functions $C_{nmn'm'}(\tau)=\left<V_{nm}(\tau)V_{n'm'}(0)\right>$.

\subsection{Correlation functions}\label{Sec:corr}

To proceed, we make the white noise assumption, where it is assumed that the bath relaxes on a time scale that is short compared to the exciton dynamics. The time dependence of the bath correlation functions is taken as a $\delta$-function,

\begin{equation}C_{nmn'm'}(\tau)\equiv
\left<V_{nm}(\tau)V_{n'm'}(0)\right>=\gamma_{nmn'm'}\delta(\tau).\end{equation}
Substitution into the time evolution equation of the density matrix, Eq. \ref{Liouville}, yields

\begin{equation}\label{timeevolution}\frac{d\rho_{nm}(t)}{dt}=-i\mathcal{L}_{sys}\rho_{nm}(t)+\sum_{n'm'}\left[\gamma_{nn'm'm}\rho_{n'm'}(t)+\gamma_{m'mnn'}\rho_{n'm'}(t)-\gamma_{nn'n'm'}\rho_{m'm}(t)-\gamma_{n'mm'n'}\rho_{nm'}(t)\right].\end{equation}

At this stage, it is worthwhile to consider the correlation matrix $\hat{\gamma}$ in more detail. First of all, the elements $\gamma_{nnnn}$ and $\gamma_{nmnm}$($n\neq m$) are simply the variances of respectively the transition energy fluctuations and the interaction fluctuations. All other elements of $\hat{\gamma}$ correspond to correlations between different fluctuations. Secondly, the correlation matrix fulfills a number of symmetries, due to the requirement of a Hermitian Hamiltonian and a set of trivial index permutation invariances,
\begin{equation}\gamma_{abcd}=\gamma_{cdab}=\gamma_{bacd}=\gamma_{abdc}.\end{equation}
We observe that there are generally three types of correlations, namely between fluctuations in \begin{itemize} \item excitation energies ($a=b,c=d$), \item excitation energies and interactions ($a=b, c\neq d$ or vice versa), \item interactions ($a\neq b,c\neq d$).\end{itemize}
A number of studies have included the first type of correlation, i.e. spatially correlated transition energy fluctuations, which have been shown to lead to appreciable changes in the transfer rates and efficiency of light-harvesting systems.~\cite{Adolphs06, Caruso09, Wu10, Strumpfer11} However, the second type of correlation has only been studied in the dimer,~\cite{Chen10} while the third type of correlation has to the best of our knowledge not been included at all in studies of EET in photosynthetic complexes. The central concern of this study is to provide an understanding of the relevance of such correlations to the time evolution of the various density matrix elements and the overall quantum efficiency.

Besides the above symmetry considerations, additional restrictions apply to the various correlations in order to describe a physical system.~\cite{Chen10} This becomes clear when one considers the expectation value of the difference between two fluctuation elements squared, which should obviously give a positive result, \begin{equation}\left<\left(V_{nm}-V_{n'm'}\right)^2\right>\propto \gamma_{nmnm}+\gamma_{n'm'n'm'}-2\gamma_{nmn'm'}\geq0.\end{equation} By considering such inequalities for various values of the indices, one finds a number of consistency conditions that need to be fulfilled,
\begin{align}\label{consistency}
\gamma_{nmnm}\geq 0\\
2\left|\gamma_{nmnm'}\right|\leq\gamma_{nmnm}+\gamma_{nm'nm'},\end{align}
where the indices $n$,$m$ and $m'$ are allowed to be equal. In addition, the magnitude of the cross-correlation between two fluctuating matrix elements is limited by the magnitude of the original fluctuations. It is straightforward to show that \begin{equation}\gamma_{nmn'm'}\leq \sqrt{\gamma_{nmnm}\gamma_{n'm'n'm'}},\end{equation} where the equality holds when the fluctuations of the Hamiltonian matrix elements $V_{nm}$ and $V_{n'm'}$ are fully (anti-)correlated. These limitations on the allowed fluctuation correlations should be kept in mind when considering the effect of including the various additional correlations.

\subsection{The Fenna-Matthews-Olson complex and light harvesting efficiency}\label{Sec:FMO}

The model light harvesting system for which we want to quantitatively probe the effects of correlations in the environmentally induced fluctuations is the Fenna-Matthews-Olson (FMO) complex. FMO is the photosynthetic complex of the green sulfur bacterium Chlorobium Tepidum, consisting of three weakly coupled, identical subunits of seven bacteriochlorophyll-a (BChl) molecules each.\cite{Ermler94, Li97, Vulto98, Cho05, Brixner05, Adolphs06, Engel07, Lee07, Panit10}  Within such a subunit, after absorption of the incoming light, the excitation energy is transferred to the reaction center where charge separation can take place. Due to their proximity to the primary light harvesting antennae, energy typically flows into the system at BChl's 1 and 6.~\cite{Li97}

We model a subunit of the FMO complex by the following experimental Hamiltonian in the site basis (in units of cm$^{-1}$),\cite{Vulto98, Cho05}

\begin{equation}\label{hamFMO} H= \left(\begin{array}{ccccccc}
280 & -106 & 8 & -5 & 6 & -8 & -4 \\
-106 & 420 & 28 & 6 & 2 & 13 & 1 \\
8 & 28 & 0 & -62 & -1 & -9 & 17 \\
-5 & 6 & -62 & 175 & -70 & -19 & -57 \\
6 & 2 & -1 & -70 & 320 & 40 & -2 \\
-8 & 13 & -9 & -19 & 40 & 360 & 32 \\
-4 & 1 & 17 & -57 & -2 & 32 & 260
\end{array} \right). \end{equation}

The corresponding exciton states, which we label in order of increasing energy, are given in Appendix \ref{appendixA}. Of particular relevance for our numerical results are the exciton states $s=3$ and $s=7$ with energies of respectively $E_3=224$ cm$^{-1}$ and $E_7=480$ cm$^{-1}$, which are the exciton states mostly associated with BChl's 1 and 2, and the lowest energy exciton state $s=1$ with energy $E_1=-24$ cm$^{-1}$ which is localized mostly on BChl 3, from where energy will be trapped to the reaction center.

In order to describe the influence of the various correlations on the light harvesting efficiency, we introduce two competing decay channels.~\cite{Ritz01, Cao08, Wu10} Exciton decay, leading to an irreversible loss of the absorbed energy, is described by adding a decay term \begin{equation}\mathcal{L}^{(dec)}_{nm}=\left(k^{(d)}_n+k^{(d)}_m\right)/2,\end{equation} where $k^{(d)}_n$ is the exciton decay rate at chromophore $n$. In the FMO complex, energy is transferred to the reaction center, and we model capture of the excitation by the trap through
a trapping term \begin{equation}\mathcal{L}^{(trap)}_{nm}=\left(k^{(t)}_n+k^{(t)}_m\right)/2,\end{equation} where $k^{(t)}_n$ is the exciton trapping rate at chromophore $n$. In the FMO complex, the reaction center is located close to BChl 3, and we will only include trapping from that particular chromophore, $k^{(t)}_n=k^{(t)}\delta_{n,3}$.

The efficiency of the process can be defined as the branching ratio between energy trapped at the reaction center and energy lost through exciton decay,~\cite{Ritz01, Cao08, Wu10}
\begin{equation} q= \frac{\sum_n k^{(t)}_n\tau_n}{\sum_n k^{(t)}_n\tau_n+\sum_n k^{(d)}_n\tau_n},
\end{equation}
where $\tau_n=\int_0^{\infty}dt \rho_{nn}(t)$ is the mean residence time at site $n$.

\section{Numerical results for the FMO complex}\label{Sec: numerical}

\begin{figure}[ht]
\begin{center}
\includegraphics[width=0.5\columnwidth]{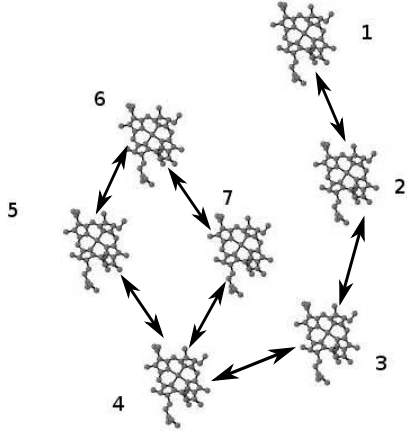}
\end{center}
    \caption{A schematic view of the FMO complex, with the strongest inter-BChl interactions denoted by arrows. The excitation will typically enter the complex at either BChl 1 or BChl 6, and will flow towards the reaction center which is most closely associated with BChl 3.
    }
\label{fig:fmoschema}
\end{figure}

As stated before, the FMO complex consists of seven coupled BChl's. In the following sections, we will first discuss the effects of uncorrelated interaction fluctuations, and subsequently proceed with a study of the effect of correlations between the various possible fluctuations. Since many of the interactions are already small to begin with, we only consider fluctuations in the strongest interactions. Likewise, we only consider correlations involving the strongest interactions, i.e. interactions between BChl's of a magnitude exceeding $25$ cm$^{-1}$, which are shown in Fig. \ref{fig:fmoschema}. In addition, we only include correlations between site energy fluctuations of site $n$ and interactions involving that same site, and similarly, correlations between interaction fluctuations that have one site in common. This choice is motivated by the fact that correlations between fluctuations are predominantly caused by shared local environments.

The numerical calculations reported in the upcoming sections concern energy flowing into the FMO complex at BChl 1. As we label the exciton states $s$ in order of increasing energy, this corresponds to initially exciting exciton states $s=3$ and $s=7$. One of our goals is to investigate the effect of including energy-interaction fluctuation correlations on the experimentally observed oscillations of the exciton populations.\cite{Engel07} Therefore, we choose an initial excitation where exciton coherence is already initially present, $\left|\psi(0)\right>=\left(\left|s=3\right>+e^{i\theta}\left|s=7\right> \right)/\sqrt{2}$. Obviously, different values of the mixing angle $\theta$ correspond to different initial populations $P_n$ of BChl's 1 and 2. In particular, $\theta=0$ implies $P_1=0.08$ and $P_2=0.89$, $\theta=\pi/2$ yields $P_1\approx P_2=0.49$, and for $\theta=\pi$ we have $P_1=0.90$ and $P_2=0.08$.

Additional calculations have been performed for energy flowing in from the other side, that is, from BChl 6. This corresponds (mostly) to initially exciting exciton states $s=5$ and $s=6$. The results do not change qualitatively, except for a considerable decrease in population oscillation frequency due to the smaller energy difference between exciton states $s=5$ and $s=6$ as compared to exciton states $s=3$ and $s=7$. For future reference, the exciton that most strongly overlaps with the trap state BChl 3 is exciton state $s=1$.

The parameters considered in our simulations have previously been used in Ref. \onlinecite{Wu10} to model the FMO system. We take all exciton decay rates equal at a value $k^{(d)}_n=k^{(d)}=1$ ns$^{-1}$, while only trapping from BChl 3 occurs, with a trapping rate $k^{(t)}=1$ ps$^{-1}$. Furthermore, we take $\gamma_0=\gamma_{nnnn}=94$ cm$^{-1}$ for the transition energy fluctuations; note that there is a factor of 2 difference in the definition of $\gamma_{nnnn}$ as compared to the quantity $\Gamma$ in Ref. \onlinecite{Wu10}.

\subsection{Uncorrelated interaction fluctuations} \label{Sec:uncorrelated fluctuations}

\begin{figure}[ht]
\begin{center}
\includegraphics[width=0.8\columnwidth]{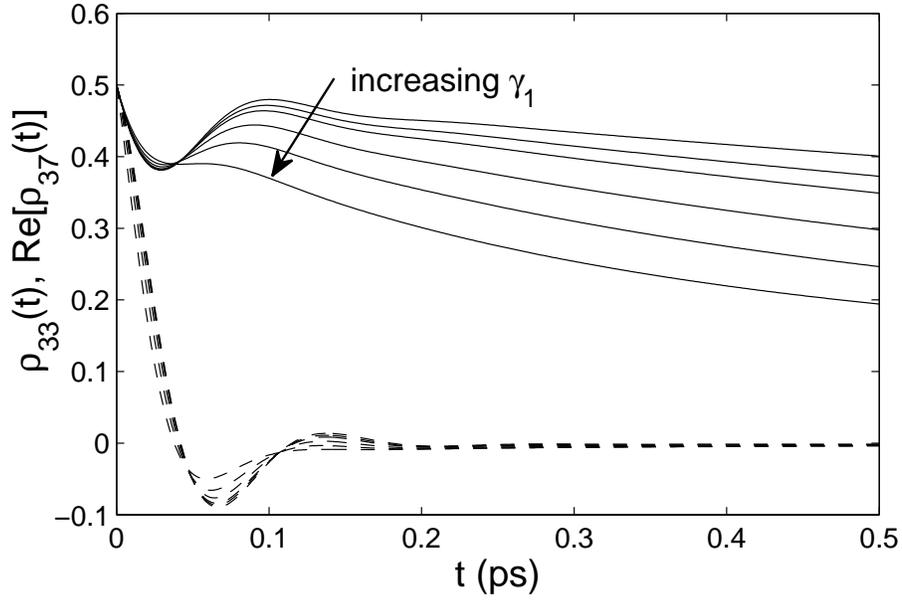}
\end{center}
    \caption{Time evolution of the exciton population $\rho_{33}$ (solid lines) and the real part of the exciton coherence $\rho_{37}$ (dashed lines), with initial phase $\theta=0$, for $\gamma_0=94$ cm$^{-1}$ and $\gamma_1=0,1,2,5,10,20$ cm$^{-1}$. Note the appreciable increase in both transfer rate and decoherence rate with $\gamma_1$.
    }
\label{fig:gamma1}
\end{figure}

The effect of interaction fluctuations can be investigated straightforwardly with the current formalism, by introducing nonzero values for the fluctuations of the off-diagonal Hamiltonian matrix elements, $\gamma_{nmnm}\equiv \gamma_1$. As stated in the previous section, only fluctuations of the strongest interactions, shown in Fig. \ref{fig:fmoschema}, are included. Note that, since $\gamma_{nmnm}$ is the expectation value of a squared quantity, it is necessarily positive. In the original works by Haken, Strobl and Reineker,\cite{Haken72, Haken73} interaction fluctuations were already considered, and also a recent paper by Chen and Silbey~\cite{Chen10} showed that interaction fluctuations can have a pronounced effect on exciton dynamics in the dimer. This makes it all the more surprising that this effect is commonly neglected in recent studies concerning the FMO complex. The time evolution of the population of exciton state $s=3$ and the (real part of the) coherence between exciton states $s=3$ and $s=7$ is shown in Fig. \ref{fig:gamma1}, clearly showing that their time evolution strongly depends on the presence of interaction fluctuations.

Analysis of the temporal evolution of the exciton populations and coherences, in particular the ones shown in Fig. \ref{fig:gamma1} and estimating the rates by assuming simple exponential decay and looking at times where the oscillations have died out, reveals that the population transfer rate and decoherence rate both increase approximately linearly with the interaction fluctuations $\gamma_1$. This is to be expected from the general form of the time evolution equations, Eq. \ref{timeevolution}. While these full equations contain a very large amount of terms, all couplings between the density matrix elements are proportional to either $\gamma_0$ or $\gamma_1$. The population transfer rate, in particular, contains only a small contribution from $\gamma_0$ terms for these parameters, and therefore the population transfer is increased dramatically upon introduction of interaction fluctuations. Note that this behavior is consistent with previous results for the dimer,~\cite{Wertheimer82, Chen10} where the exciton population transfer rate for a strongly detuned dimer is to a very large extent dominated by the interaction fluctuations. This is analogous to the situation in the FMO complex, where the differences between the BChl transition energies are also typically large compared to the interactions. More specifically, for a dimer detuned by an energy difference $\Delta E$ and having an interaction strength $J$, the mixing angle $\theta^{\prime}$ defined by $\tan \theta^{\prime}=2J/\Delta E$ is small. It can analytically be shown~\cite{Wertheimer82, Chen10} that the population transfer rate $\Gamma$ between the two exciton states is given by $\Gamma=\gamma_0 \sin^2\theta^{\prime}+2\gamma_1\cos^2\theta^{\prime}$. However, since the first term is typically very small for strongly detuned dimers (i.e., $\Delta E \gg J$), the population transfer rate is fully dominated by the second term as soon as interaction fluctuations are present. An analysis of the population evolution shown in Fig. \ref{fig:gamma1} confirms that the population transfer rates in FMO are indeed also proportional to $\gamma_1$. Thus, interaction fluctuations lead to dramatic increases in population transfer rates, in turn enhancing the efficiency of the FMO complex.

\begin{figure}[ht]
\begin{center}
\includegraphics[width=0.8\columnwidth]{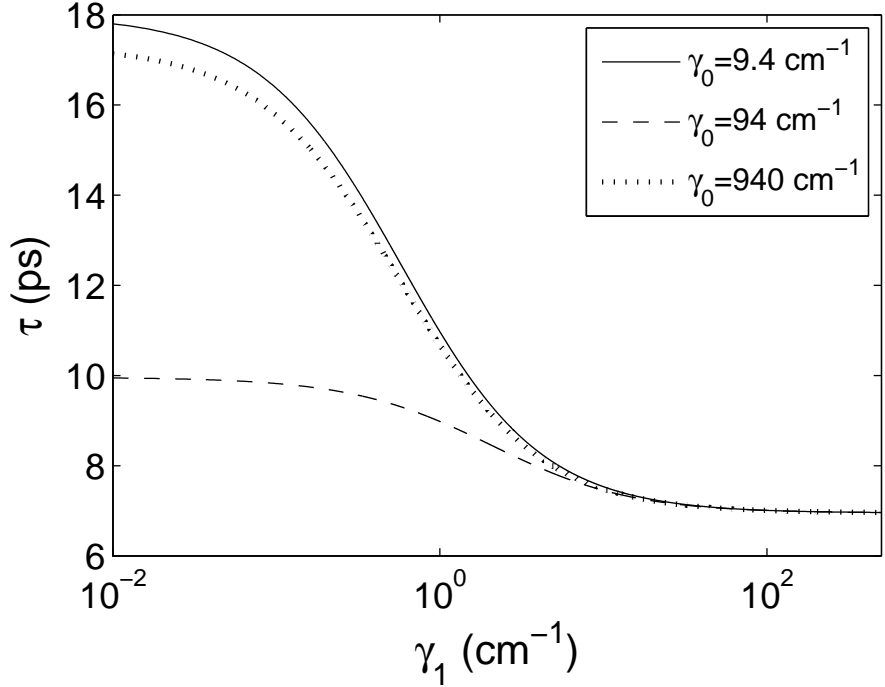}
\end{center}
    \caption{Dependence of the harvesting time on the interaction fluctuation magnitude $\gamma_1$, for various values of the energy fluctuations $\gamma_0$. Notice the decrease in harvesting time with $\gamma_1$ as a result of increasing population transfer rates.
    }
\label{fig:gamma1-efficiency}
\end{figure}

To further illustrate the increase in population transfer rates with increasing interaction fluctuations, Fig. \ref{fig:gamma1-efficiency} shows the harvesting time as a function of the interaction fluctuation amplitude $\gamma_1$. In Fig. \ref{fig:gamma1-efficiency}, we consider the energy fluctuation magnitude $\gamma_0$ that optimizes the harvesting time, $\gamma_0=94$ cm$^{-1}$, and an increase respectively decrease of one order of magnitude. Note that increasing $\gamma_1$ leads to more efficient light harvesting due to a corresponding increase in population transfer within the FMO complex; this holds for different values of the energy fluctuations $\gamma_0$ and for different initial conditions. Additionally, while the optimal dephasing value $\gamma_0$ does not change with $\gamma_1$, the optimum becomes considerably less deep, making the FMO complex more robust to variations in the dephasing rate $\gamma_0$. For very large, in fact unphysically large, interaction fluctuations, the efficiency of the light harvesting complex becomes independent of the energy fluctuation magnitude. This situation corresponds to large population transfer rates, dominated by the interaction fluctuations $\gamma_1$, and subsequent rapid redistribution of population over all BChl's. The harvesting time will in that case converge to $\tau=7$ ps, corresponding to a trapping rate $k_t=1$ ps$^{-1}$ from BChl 3 where at all times 1/7th of the total untrapped population will reside.

\subsection{Effect of correlated energy-interaction fluctuations} \label{Sec:EJfluc}

\subsubsection{Parameter set 1} \label{subsec: ps1}

Here, we consider $\gamma_0=\gamma_{nnnn}=94$ cm$^{-1}$ which corresponds to parameters close to the optimal energy transport conditions reported in Ref. \onlinecite{Wu10}. The magnitude of the allowed correlations is limited by Eqs. \ref{consistency}; to allow for appreciable amounts of correlation, we take a large interaction fluctuation value $\gamma_1=\gamma_{nmnm}=40$ cm$^{-1}$ ($n \neq m$), with the same provisions as before. In Sec. \ref{subsec: ps2}, we discuss the effect of correlated energy and interaction fluctuations for a smaller, more realistic value of the interaction fluctuation magnitude $\gamma_1$. As stated in Sec. \ref{Sec: numerical}, we only include correlations that involve the strongest interactions (see Fig. \ref{fig:fmoschema}) and we focus on correlations between interaction fluctuations and energy fluctuations of the BChl's involved in that particular interaction; $\gamma_{n'n'nm}\propto\left(\delta_{n'n}+\delta_{n'm}\right)$, $n \neq m$.

First of all, taking all energy-interaction fluctuation correlations equal in sign and magnitude leads to negligible effects. This is consistent with a previous study by Chen and Silbey, where it was shown that in a dimer the relevant quantity is the difference between the two transition energy-interaction fluctuation correlations.\cite{Chen10} While the full time evolution equations for the more complicated FMO system are more involved, a similar result is observed to hold. This is not surprising, as fluctuation correlations of equal sign would imply that a change of interaction would mean that both energies involved change in the same direction, leaving the energy difference unchanged. Therefore, we choose the various energy-interaction correlations of equal magnitude, but with alternating sign; that is, we take $\gamma_{1112}=-\gamma_{2212}=\gamma_{2223}=-\gamma_{3323}\equiv \gamma_2$ and so on. There is an ambiguity near BChl 4, as it strongly interacts with three other BChl's and it is therefore necessary to choose two out of $\gamma_{4434}$, $\gamma_{4454}$ and $\gamma_{4474}$ with equal sign. We take $\gamma_{4434}=\gamma_{4474}=-\gamma_{4454}=-\gamma_2$; other choices result only in inconsequential numerical changes.

\begin{figure}[ht]
\begin{center}
\includegraphics[width=0.8\columnwidth]{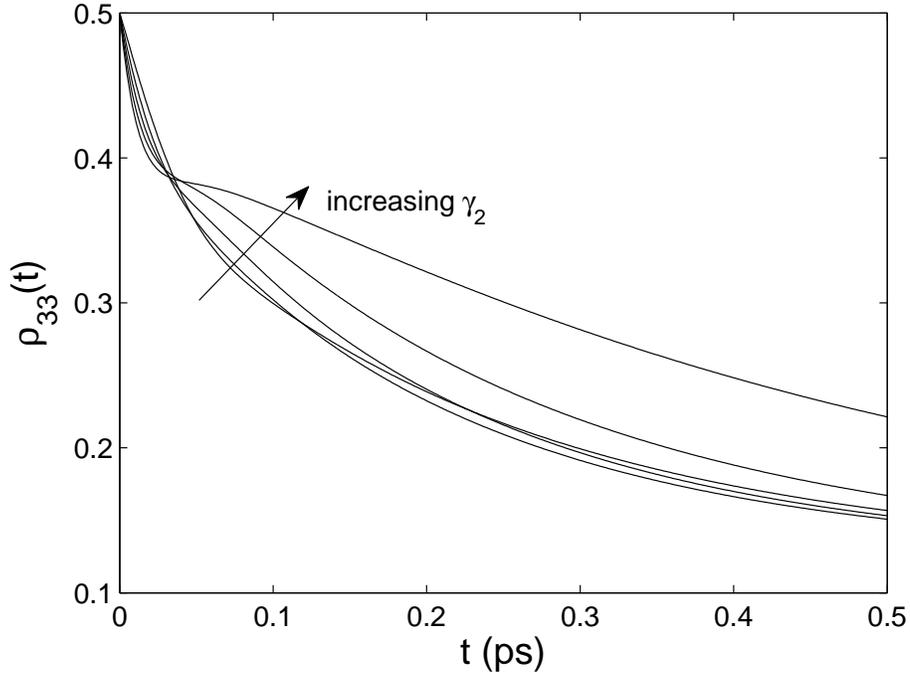}
\end{center}
    \caption{Time evolution of the exciton population $\rho_{33}$, $\theta=0$, with $\gamma_0=94$ cm$^{-1}$ and $\gamma_1=40$ cm$^{-1}$. The considered correlation values are $\gamma_2=-40,-20,0,20,40$ cm$^{-1}$.
    }
\label{fig:33EJ}
\end{figure}

A few observations can be made regarding these plots. First of all, energy-interaction fluctuation correlations affect both the energy transfer rates and the coherence decay rates. Depending on the sign of the correlations, both increases and decreases of the various rates can occur. This can be observed in Fig. \ref{fig:33EJ}, where the population of exciton state $s=3$ shows appreciable differences in transfer rate to other exciton states for different correlation magnitudes. While this particular exciton state and initial condition shows an increase in the transfer rate upon increasing correlations, this is not general; a decrease in transfer rate may be obtained for different exciton states and initial conditions. Secondly, it is clear that these parameters lead to a quick redistribution of population over all chromophores, only showing oscillatory features on a short time scale. Therefore, in the next subsection, we consider the same quantities for different parameters, where the effect on population oscillations can be observed more clearly.

\begin{figure}[ht]
\begin{center}
\includegraphics[width=0.8\columnwidth]{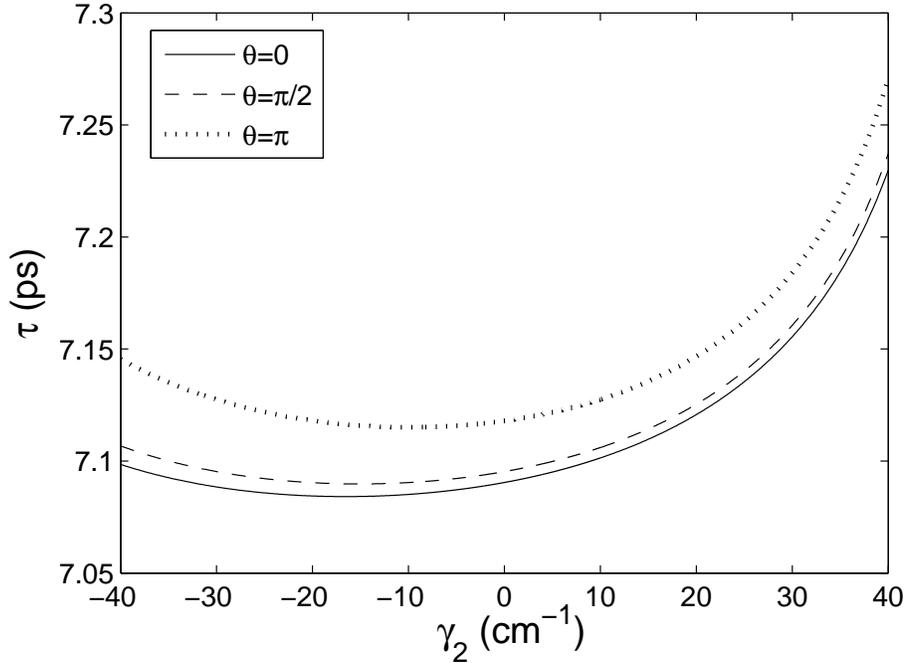}
\end{center}
    \caption{Dependence of the harvesting time on the magnitude of the energy-interaction fluctuation correlations, for various initial conditions and with $\gamma_0=94$ cm$^{-1}$ and $\gamma_1=40$ cm$^{-1}$. Note that negative values of $\gamma_2$ are required for enhancement of the efficiency, and that the optimal correlation value depends on the initial condition.
    }
\label{fig:eff1}
\end{figure}

It is also possible to investigate the dependence of the harvesting efficiency, here quantified through the average trapping time $\tau$, on the magnitude of the energy-interaction fluctuation correlation $\gamma_2$. This is shown in Fig. \ref{fig:eff1}; the effects are rather limited, with changes of at most a few percent. Both the overall effect of accounting for these correlations and the dependence on the initial condition, defined by the mixing angle $\theta$, is rather weak. This is related to the fact that we are close to the conditions that constitute optimal harvesting behavior (i.e., $\gamma_0=94$ cm$^{-1}$), which is corroborated by the harvesting times which are only slightly above $\tau=7$ ps. In other words, the bottleneck in the overall harvesting process is not population transfer between the BChl's, but the trapping of excitations from BChl 3 into the reaction center. Therefore, the effect of limited changes in the population transfer rates have relatively little effect on the overall efficiency. Note that negative values of $\gamma_2$ may give a small enhancement of the harvesting time; depending on the initial conditions, a nonzero optimal choice for the correlations may be made such that light harvesting is at its fastest.

Again, we note that the fluctuations considered here lead to a rapid decay of the coherences. In the next section, we will consider parameters that lead to slower decoherence and longer lived population oscillations, in order to evaluate the effect of energy-interaction fluctuation correlations on population oscillations.

\subsubsection{Parameter set 2} \label{subsec: ps2}

\begin{figure}[ht]
\begin{center}
\includegraphics[width=0.8\columnwidth]{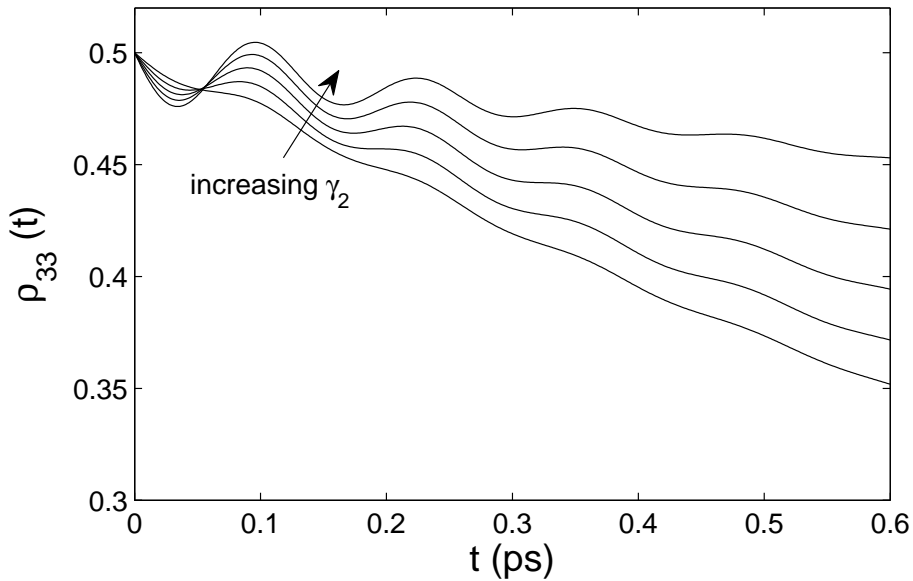}
\end{center}
    \caption{Time evolution of the exciton population $\rho_{33}$, $\theta=0$ and with $\gamma_0=10$ cm$^{-1}$ and $\gamma_1=2$ cm$^{-1}$. We consider the values $\gamma_2=-4,-2,0,2,4$ cm$^{-1}$.
    }
\label{fig:ps2}
\end{figure}

In order to clearly observe the effect of additional energy-interaction fluctuation correlations on population oscillations, it is necessary to switch to parameters where the decoherence occurs on a slower timescale. Here, we consider $\gamma_0=\gamma_{nnnn}=10$ cm$^{-1}$, and $\gamma_1=\gamma_{nmnm}=2$ cm$^{-1}$ ($n \neq m$) for the interactions shown in Fig. \ref{fig:fmoschema}. The same observations as before hold with regards to the signs of the correlations: alternating signs are required for observable effects and are used throughout this section, while a negative $\gamma_2$ can produce enhancement of the transport efficiency. Again, as an example we consider the time evolution of the population of the exciton states $s=3$ and initial phase $\theta=0$ for various values of $\gamma_2$, shown in Fig. \ref{fig:ps2}.

It can now be clearly seen that the populations oscillate in time, an effect which is induced by the coherence that is present between the two exciton states. Indeed, the period of the oscillations corresponds to the energy difference between the exciton states, and is identical to the coherence oscillation period, which corresponds to $\tau_{osc}=130$ fs. A clear augmentation of the amplitude of the oscillations can be observed when one increases the magnitude of $\gamma_2$. The introduction of energy-interaction fluctuation correlations can thus enhance the population oscillations, which are driven by the presence of coherence. As before, we also observe a change in population transfer rates, with in particular a strong suppression of population transfer with increasing $\gamma_2$.

\begin{figure}[ht]
\begin{center}
\includegraphics[width=0.8\columnwidth]{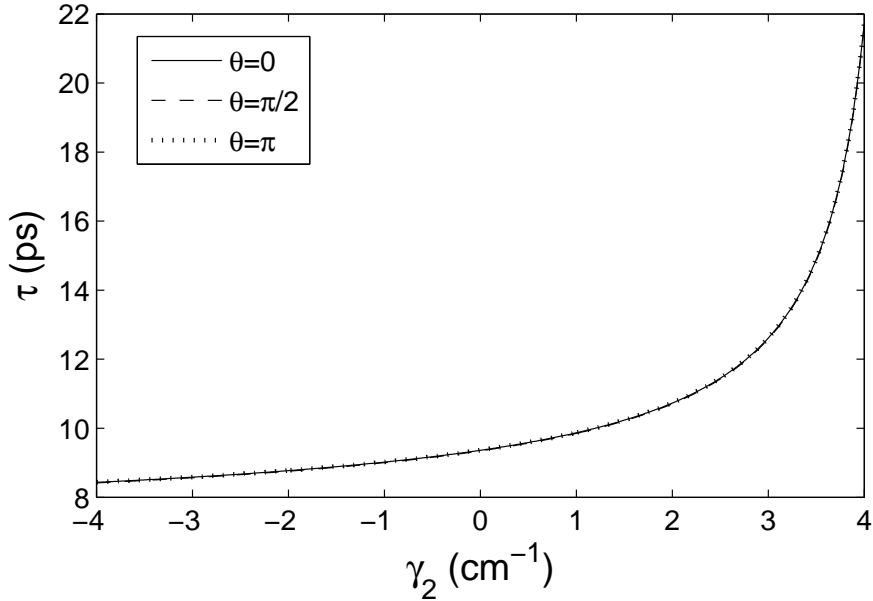}
\end{center}
    \caption{Dependence of the harvesting time on the magnitude of the energy-interaction fluctuation correlations, for various initial conditions and with $\gamma_0=10$ cm$^{-1}$ and $\gamma_1=2$ cm$^{-1}$. Note that negative values of $\gamma_2$ are required for enhancement of the efficiency.
    }
\label{fig:eff2}
\end{figure}

The dependence of the harvesting efficiency on the correlation magnitudes can be calculated exactly as before, and is shown in Fig. \ref{fig:eff2}. There is now a stronger dependence on the magnitude of the correlations, which is caused by the fact that this parameter set constitutes suboptimal conditions, so that in this case changes in the population transfer rates have a larger effect on the overall harvesting time. Also, the behavior is almost independent of the initial conditions, even more so than for the parameters in the previous section. Negative values for $\gamma_2$ are required for enhancement, where the transport is now optimized by choosing $\gamma_2$ as negative as possible. Positive values of $\gamma_2$ lead to a strong increase in harvesting time; this corresponds to the strong decrease in the relevant population transfer rates, as observed in for example Fig. \ref{fig:ps2}.

\subsection{Effect of correlated interaction-interaction fluctuations}\label{Sec:JJfluc}

\begin{figure}[ht]
\begin{center}
\includegraphics[width=0.8\columnwidth]{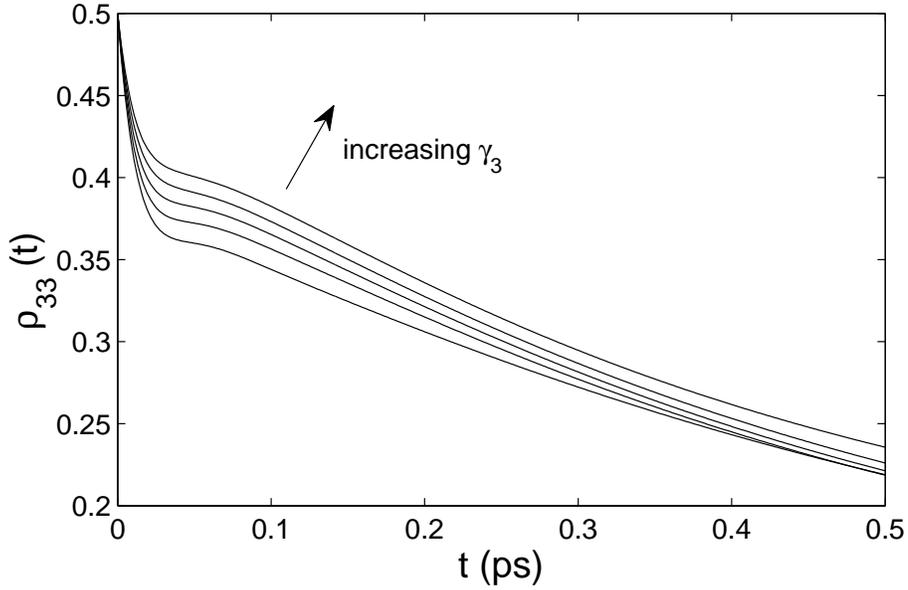}
\end{center}
    \caption{Time evolution of the exciton population $\rho_{33}$, $\theta=0$, $\gamma_0=94$ cm$^{-1}$, $\gamma_1=40$ cm$^{-1}$, and $\gamma_2=40$ cm$^{-1}$. The considered correlation values are $\gamma_3=-30,-15,0,15,30$ cm$^{-1}$.
    }
\label{fig:33JJ}
\end{figure}

It is straightforward to include correlations between the various interaction fluctuations. As in the previous section, we anticipate that such correlations are strongest between interactions that have one molecule in common, and in addition we focus on the strongest interactions. First of all, it is important to note that while such correlations may very well be present, also these are naturally limited in magnitude in order to fulfill the consistency conditions Eqs. \ref{consistency}. In particular, we have

\begin{equation}
\left<\left(V_{ab}-V_{ac}\right)^2\right>\propto \gamma_{abab}+\gamma_{acac}-2\gamma_{abac}>0,
\end{equation}
where $a \neq b, a\neq c$. From this, it follows that
\begin{equation}
\left|\gamma_{abac}\right| \leq \frac{1}{2}\left(\gamma_{abab}+\gamma_{acac}\right).\end{equation}

\begin{figure}[ht]
\begin{center}
\includegraphics[width=0.7\columnwidth]{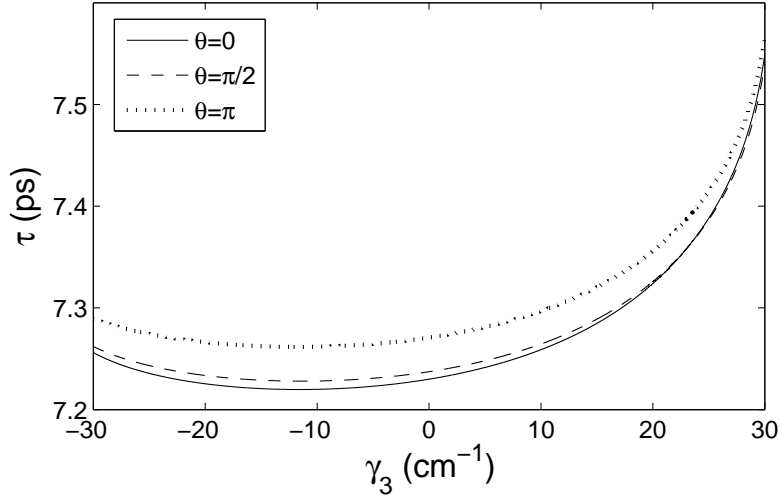}
\end{center}
    \caption{Dependence of the harvesting time on the magnitude $\gamma_3$ of the interaction-interaction fluctuation correlations, for various initial conditions. We have chosen $\gamma_0=94$ cm$^{-1}$, $\gamma_1=40$ cm$^{-1}$ and $\gamma_2$=40 cm$^{-1}$. The different offsets of the curves are consistent with the observed $\theta$-dependence of the harvesting time in Fig. \ref{fig:eff1}.
    }
\label{fig:qC}
\end{figure}

Nevertheless, it is instructive to quantify the effects of these additional correlations. The conclusions are typically similar to those in the previous section, although the effects are typically less pronounced. For brevity, we will not show all the same figures as before, but we show a typical result in Fig. \ref{fig:33JJ}. Here, we again consider the parameters previously used in Sec. \ref{subsec: ps1}, with in addition an energy-interaction fluctuation correlation magnitude $\gamma_2=40$ cm$^{-1}$ with the sign conventions defined there. First of all, here too alternating signs are required to amplify the effect of interaction-interaction fluctuation correlations. We take the interaction-interaction fluctuation correlations of the form $\gamma_{1223}=-\gamma_{2334}=\gamma_{3445}=\gamma_3$ et cetera. Again there is some ambiguity in the definition of the signs for the correlations around BChl 4, however, since the details of those choices hardly matter for the initial conditions we have chosen, we will not go in to this in more detail. We show time evolution plots only for $\theta=0$, behavior for other parameters is not significantly different. First of all, it is clear that interaction-interaction fluctuation correlations, too, lead to changes in the population transfer and dephasing rates, as exemplified by the population evolution of exciton state 3 shown in Fig. \ref{fig:33JJ}. In addition, while there is a small change in the size of the populations, the amplitude of the oscillations is influenced by the interaction-interaction fluctuation correlation magnitude to only a small extent. These conclusions are observed to hold for different sets of parameters and initial conditions.

Fig. \ref{fig:qC} shows the dependence of the harvesting time on the magnitude of the interaction-interaction fluctuation correlations. Again, there is an optimum for some nonzero value of the correlation magnitude $\gamma_3$; the exact value of the optimal magnitude for the correlations depends on the initial conditions and on the choice for the energy-interaction correlations. The behavior for other values of the initial phase factor $\theta$ is very similar, implying that the transfer rates and other relevant parameters for trapping will only depend weakly on possible interaction-interaction fluctuation correlations. While Fig. \ref{fig:qC} shows that the effect of interaction-interaction fluctuation correlations is roughly of the same magnitude as the effect of transition energy-interaction fluctuation correlations, this is only the case if the former occurs in conjunction with the latter. The changes induced by interaction-interaction fluctuation correlations for the case of $\gamma_2=0$ cm$^{-1}$ are considerably smaller. Finally, as was previously the case in Sec. \ref{Sec:EJfluc}, a more suboptimal choice of parameters leads to a larger effect of changes in transfer rates, and will thus also amplify the effect of including interaction-interaction fluctuation correlations to an extent.

\section{Conclusions}\label{Sec:conclusions}

Since in principle environmental changes will induce fluctuations of both transition energies and interactions that should be correlated to some extent, we have performed a study of the effect of such fluctuations and their possible correlations on the energy transfer and excitation dynamics in photosynthetic complexes, specifically focusing on the Fenna-Matthews-Olson complex as a model system. The inclusion of interaction fluctuations and correlations between transition energy fluctuations and interaction fluctuations, and between different interaction fluctuations, has been studied by a straightforward generalization of the Haken-Strobl-Reineker model. These additional correlations have been shown to be naturally limited in magnitude, in order to correspond to a physically consistent system.

Interaction fluctuations on the excitation dynamics in photosynthetic complexes can in general not be neglected. Not only will these in general occur, but it has been shown that even small values of interaction fluctuations can already lead to an appreciable increase in transfer rates and a corresponding increase in efficiency of the FMO complex. In particular, for suboptimal values of the energy fluctuations, interaction fluctuations can lead to a considerable optimization of the light harvesting process, even to such an extent that the transfer rates and the eventual efficiency is dominated by interaction fluctuations and not by energy fluctuations. The presence of interaction fluctuations may thus make the efficiency of the FMO complex more robust to changes in the dephasing rate. Generally, for various initial conditions and energy fluctuation magnitudes, an increase in the efficiency of the FMO complex is observed with increasing interaction fluctuation amplitude.

When including correlated transition energy fluctuations and interaction fluctuations, it is first of all possible to observe somewhat enhanced population oscillations, driven by the coherence between the initially excited exciton states. Depending on the signs of the various correlations, a suppression of the oscillations can also occur. In addition, one sees that the transfer rates between exciton states is modified, which in turn leads to accompanying changes in the overall efficiency of the photosynthetic complex. The overall effects on the efficiency are limited in scope when one considers parameter values that approximately correspond to an optimal functioning of the light harvesting complex. In that case, not the excitation transfer but the trapping into the reaction center is the bottleneck, so that changes in transfer rates will only change the harvesting time by up to a few percent. For less optimal conditions, the effects can be considerably larger, as the excitation transfer processes play a larger role in determining the overall harvesting time. Typically, a small increase in the light harvesting efficiency can be obtained by using a nonzero amount of correlation between the aforementioned fluctuations. By the same approach, one can quantify how correlations between the various interaction fluctuations modify previously obtained results. The effect of these additional correlations turns out to be qualitatively very similar to that of transition energy-interaction fluctuation correlations, and are quantitatively of a comparable magnitude. The net effect of such correlations is limited to small changes in the scattering rate and resultant changes in the overall efficiency.

This study thus suggests that one should not neglect interaction fluctuations in models describing the energy transfer in photosynthetic complexes. Correlations may also play a relevant role in the excitation dynamics, and this may (but does not necessarily) translate into appreciable changes in the efficiency.

\section*{Acknowledgments}
The authors thank Dr. Jianlan Wu, Dr. Xin Chen and Dr. Jianshu Cao for useful discussions.  This work is supported by ARO under grant W911NF-09-0480, and DARPA grant N66001-10-1-4063.

\appendix

\section{Exciton states and energies in the FMO complex}\label{appendixA}

The exciton states are found by diagonalization of the Hamiltonian matrix (\ref{hamFMO}); their coefficients and energies are given in the table below.

\begin{center}
    \begin{tabular}{| r | r r r r  r  r  r | r |}
    \hline
     & ~BChl 1 & ~BChl 2 & ~BChl 3 & ~BChl 4 & ~BChl 5 & ~BChl 6 & ~BChl 7 & $E_s$ (cm$^{-1}$)  \\ \hline
    s=1 & 0.046 & 0.076 & -0.940 & -0.321 & -0.066 & -0.032 & -0.005 &  -24 \\ 
    s=2 & -0.066 & -0.036 & 0.285 & -0.786 & -0.319 & 0.064 & -0.435 &  139 \\
    s=3 & 0.877 & 0.461 & 0.089 & -0.011 & -0.090 & 0.043 & -0.019 &  224 \\ 
    s=4 & 0.013 & 0.000 & -0.138 & 0.340 & 0.254 & 0.268 & -0.854 &  276 \\ 
    s=5 & 0.040 & 0.098 & 0.073 & -0.265 & 0.712 & -0.629 & -0.103 &  311 \\ 
    s=6 & -0.080 & 0.111 & -0.034 & 0.304 & -0.560 & -0.710 & -0.264 &  408 \\ 
    s=7 & -0.465 & 0.871 & 0.043 & -0.007 & 0.032 & 0.144 & 0.038 &  480 \\ \hline
    \end{tabular}
\end{center}

\end{document}